\documentstyle[aps]{revtex}

\begin{document}

\title{Dependence of the MHD shock thickness on the finite electrical conductivity}

\author{Alejandra Kandus and Reuven Opher}

\address{Departamento de Astronom\'{\i}a, IAG-USP, Rua do Mat\~{a}o 1226, 
\\
Cidade Universitaria, CEP: 05508-900,
S\~{a}o Paulo, SP, Brazil.}

\maketitle

\begin{abstract}
The results of MHD\ plane shock waves with infinite electrical conductivity
are generalized for a plasma with a finite conductivity. We derive the
adiabatic curves that describe the evolution of the shocked gas as well as
the change in the entropy density. For a parallel shock (i.e., in which the
magnetic field is parallel to the normal to the shock front) we find an
expression for the shock thickness which is a function of the ambient
magnetic field and of the finite electrical conductivity of the plasma. We
give numerical estimates of the physical parameters for which the shock
thickness is of the order of, or greater than, the mean free path of the
plasma particles in a strongly magnetized plasma.
\end{abstract}

\section{Introduction}

Previously, shock waves in plasmas, both relativistic and non-relativistic,
were studied assuming ideal MHD, \cite{ecm-land,priest,anile}. Although this
theory is suitable for studying most astrophysical shock waves, such as
those in hot rarified astrophysical plasmas, where the electrical
conductivity is extremely high and the magnetic field is weak, it is
interesting to study the effect of the simplest dissipative process in
non-ideal MHD, that due to a finite value of the electrical conductivity 
$\sigma $. We generalize here the results for planar shock waves to
non-ideal, non-relativistic MHD. The junction conditions that must be
satisfied across a shock wave in a non-ideal plasma are given in Sec. II. We
then derive the adiabatic curve with corrections due to $\sigma $ and find
the corresponding change in the entropy density across the shock. We perform
this analysis for an oblique shock (i.e., in which $\vec{B}$ is neither
parallel nor perpendicular to the normal to the shock surface) as well as
for a parallel one (i.e., in which $\vec{B}$ is parallel to the normal to
the shock surface). In Sec. III, we find a closed expression for the shock
thickness for a parallel shock in a strongly magnetized plasma, and estimate
its value for some physical situations. Our conclusions are discussed in
Sec. IV.

\section{Junction conditions and adiabatic curve}

For non ideal MHD, the fluxes of mass, energy, and momentum are given by
Eqs. (\ref{OO}-\ref{c}), respectively, \cite{ecm-land,flm-land},

\begin{equation}
\vec{M}=\rho \vec{v},  \label{OO}
\end{equation}

\begin{equation}
\vec{q}=\rho \vec{v}\left( \frac{1}{2}v^{2}+{\sf w}\right) +\frac{1}{4\pi }
\vec{B}\times \left( \vec{v}\times \vec{B}\right) -\frac{c^{2}}{16\pi
^{2}\sigma }\vec{B}\times \left( \vec{\nabla}\times \vec{B}\right) ,
\label{a}
\end{equation}
and

\begin{equation}
\Pi _{ik}=\rho v_{i}v_{k}+p\delta _{ik}-\frac{1}{4\pi }\left( B_{i}B_{k}-
\frac{1}{2}B^{2}\delta _{ik}\right) ,  \label{b}
\end{equation}
where $\rho $ is the fluid density, $\vec{v}$ the velocity, ${\sf w}$ the
enthalpy per unit mass, $\vec{B}$ the ambient magnetic field, and $p$ is the
fluid pressure. The electric field is 
\begin{equation}
\vec{E}=\frac{c}{4\pi \sigma }\left( \vec{\nabla}\times \vec{B}\right) -
\frac{\vec{v}}{c}\times \vec{B},  \label{c}
\end{equation}
where Ohm's law in its simplest form \cite{priest,spitzer} was used.

\subsection{Junction conditions}

We assume a two dimensional, planar, shock wave in the $y-z$ plane. The
normal to the transition surface is in the $-x$ direction. The velocity
field can be decomposed into perpendicular and tangential components to the
surface of transition, $\vec{v}=\left( v_{x},\vec{v}_{t}\right) $. It is
assumed that all quantities vary as a function of $x$ . Let $n^{x}$ be a
unit vector normal to the transition surface. We then have the
hydrodynamical junction conditions, \cite{flm-land},

\begin{equation}
\left[ \rho v_{x}n^{x}\right] =0  \label{e}
\end{equation}

\begin{equation}
\left[ q_{x}n^{x}\right] =0  \label{f}
\end{equation}

\begin{equation}
\left[ \Pi _{ix}n^{x}\right] =0  \label{g}
\end{equation}
where $i$ is\ $x$ ($t$) [normal (tangential) to the shock surface] and 
$\left[ {}\right] $ means the difference between the value of the
corresponding quantities far upstream (which we denote by subscript  ``1'')
and the value at some point in the shock (no subscript). It is also assumed
that both far upstream and far downstream, all gradients vanish (i.e., the
fields and flows are uniform).

In the MHD case that we are considering, Eqs. (\ref{e}), (\ref{f}), and (\ref
{g}) must be suplemented with the electromagnetic junction conditions, i.e.,
that the normal component of the magnetic field and the tangential component
of the electric field must be constant across the shock surface:

\begin{equation}
\left[ B_{n}\right] =0,  \label{p}
\end{equation}

\begin{equation}
\left[ \vec{E}_{t}\right] =\left[ \frac{c}{4\pi \sigma }\left( \partial
_{x}\times \vec{B}\right) _{t}-\frac{v_{x}}{c}\vec{B}_{t}+\frac{B_{n}}{c}
\vec{v}_{t}\right] =0.  \label{q}
\end{equation}

Eq. (\ref{e}) states that the mass flux along $x$ is conserved, i.e., 
$\rho v_{x}=j=const$. We write $\rho =1/V$, where $V$ is the specific 
volume, and
replace $v_{x}=jV$ in the other junction conditions, obtaining

\begin{equation}
j\left[ \frac{1}{2}j^{2}V^{2}+\frac{1}{2}v_{t}^{2}+{\sf w}\right] +
\frac{1}{4\pi }j\left[ VB_{t}^{2}\right] -\frac{1}{4\pi }B_{x}\left[ 
\vec{v}_{t}.\vec{B}_{t}\right] -\frac{c^{2}}{16\pi ^{2}\sigma }\left[ 
\partial _{x}B_{t}^{2}
\right] =0,  \label{t}
\end{equation}

\begin{equation}
j^{2}\left[ V\right] +\left[ p\right] +\frac{1}{8\pi }\left[ B_{t}^{2}\right]
=0,  \label{u}
\end{equation}

\begin{equation}
j\left[ \vec{v}_{t}\right] -\frac{1}{4\pi }B_{n}\left[ \vec{B}_{t}\right] =0,
\label{v}
\end{equation}

\begin{equation}
\frac{c}{4\pi \sigma }\left[ \left( \partial _{x}\times \vec{B}_{t}\right) 
\right] -\frac{j}{c}\left[ V\vec{B}_{t}\right] +\frac{B_{n}}{c}\left[ 
\vec{v}_{t}\right] =0.  \label{w}
\end{equation}

\subsection{Adiabatic curve and entropy density change}

We derive the expression for the adiabatic curve and the corresponding
entropy density change. From Eqs. (\ref{v}) and (\ref{w}), we obtain

\begin{equation}
\frac{1}{4\pi }B_{n}^{2}\left[ \vec{B}_{t}\right] =j^{2}\left[ V\vec{B}_{t}
\right] -\frac{jc^{2}}{4\pi \sigma }\left[ \left( \partial _{x}\times 
\vec{B}_{t}\right) _{t}\right] .  \label{z}
\end{equation}
From Eq. (\ref{v}), we have $\left[ \vec{v}_{t}\right] =B_{n}\left[ \vec{B}
_{t}\right] /4\pi j$. We can therefore complete the squares in Eq. (\ref{t}
), obtaining

\begin{eqnarray}
&&\left[ {\sf w}\right] +\frac{1}{2}j^{2}\left[ V^{2}\right] +\frac{1}{2}
\left[ \left( v_{t}-\frac{1}{4\pi }\frac{B_{n}}{j}\vec{B}_{t}\right) ^{2}
\right] -\frac{1}{32\pi ^{2}}\frac{B_{n}^{2}}{j^{2}}\left[ B_{t}^{2}\right] 
+
\frac{1}{4\pi j}B_{n}\left[ \vec{v}_{t}.\vec{B}_{t}\right]   \nonumber \\
&&+\frac{1}{4\pi }\left[ VB_{t}^{2}\right] -\frac{1}{4\pi j}B_{n}\left[ 
\vec{v}_{t}.\vec{B}_{t}\right] -\frac{c^{2}}{16\pi ^{2}j\sigma }\left[ 
\partial_{x}\left( B_{t}^{2}\right) \right] \left. =\right. 0.  \label{ac}
\end{eqnarray}
From Eq. (\ref{v}), the third term of Eq. (\ref{ac}) is zero and we are left
with

\begin{equation}
\left[ {\sf w}\right] +\frac{1}{2}j^{2}\left[ V^{2}\right] -\frac{1}{32\pi
^{2}}\frac{B_{n}^{2}}{j^{2}}\left[ B_{t}^{2}\right] +\frac{1}{4\pi }\left[
VB_{t}^{2}\right] -\frac{c^{2}}{16\pi ^{2}j\sigma }\left[ \partial
_{x}\left( B_{t}^{2}\right) \right] =0.  \label{ad}
\end{equation}
Using Eq. (\ref{z}), we can write the third term of Eq. (\ref{ad}) as

\begin{equation}
\frac{1}{32\pi ^{2}}\frac{B_{n}^{2}}{j^{2}}\left[ B_{t}^{2}\right] =
\frac{1}{8\pi }\left[ VB_{t}^{2}\right] +\frac{1}{8\pi }\left( V-V_{1}\right) 
\vec{B}.\vec{B}_{1}-\frac{c^{2}}{32\pi ^{2}j\sigma }\left( \vec{\nabla}\times 
\vec{B}
\right) .\left( \vec{B}_{t}+\vec{B}_{1t}\right) .  \label{am}
\end{equation}
From momentum conservation, we obtain

\begin{equation}
j^{2}=-\frac{\left( p-p_{1}\right) }{\left( V-V_{1}\right) }-\frac{1}{8\pi }
\frac{\left( B_{t}^{2}-B_{1t}^{2}\right) }{\left( V-V_{1}\right) }.
\label{an}
\end{equation}
Using Eqs. (\ref{am}) and (\ref{an}) in Eq. (\ref{ad}) and ${\sf w}
=\varepsilon +pV$, we have

\begin{eqnarray}
&&\varepsilon -\varepsilon _{1}+\frac{1}{2}\left( p+p_{1}\right) \left(
V-V_{1}\right) +\frac{1}{16\pi }\left( \vec{B}_{t}-\vec{B}_{1t}\right)
^{2}\left( V-V_{1}\right)  \label{aq} \\
&&+\frac{c^{2}}{32\pi ^{2}j\sigma }\left[ \left( \vec{\nabla}\times \vec{B}
\right) .\left( \vec{B}_{t}+\vec{B}_{1t}\right) -2\partial _{x}\left(
B_{t}^{2}\right) \right] \left. =\right. 0.  \nonumber
\end{eqnarray}
The first four terms are found in the equation for ideal MHD\ (e.g. Ref. 
\cite{ecm-land}), while the last two are due to the finite electrical
conductivity. The first term in the square brackets is due to the fact that 
$B_{n}\neq 0$; the second is the contribution from the tangential component
of the magnetic field.

To obtain the entropy density change, we follow the procedure found in the
standard literature, \cite{flm-land} and develop $V-V_{1}$ in powers of 
$\left( p-p_{1}\right) $. We also expand $\left( {\sf w}-{\sf w}_{1}\right) $
in powers of $\left( p-p_{1}\right) $ and to first order in powers of 
$\left( {\sf s}-{\sf s}_{1}\right) $. The resulting expression is

\begin{eqnarray}
T\left( {\sf s}-{\sf s}_{1}\right)  &=&\frac{1}{12}\left( \frac{\partial
^{2}V}{\partial p^{2}}\right) _{s}\left( p-p_{1}\right) ^{3}-\frac{1}{16\pi }
\left( \frac{\partial V}{\partial p}\right) _{s}\left( B_{t}-B_{1t}\right)
^{2}\left( p-p_{1}\right)   \nonumber \\
&&-\frac{c^{2}}{32\pi ^{2}j\sigma }\left[ \left( \vec{\nabla}\times \vec{B}
\right) .\left( \vec{B}_{t}+\vec{B}_{1t}\right) -2\partial _{x}\left(
B_{t}^{2}\right) \right] .  \label{au}
\end{eqnarray}
The first two terms are found in the equation for ideal MHD and the last two
are the corrections due to a finite $\sigma $. In the following section, we
repeat the calculations for a perpendicular shock and find that in the
expressions for the adiabatic curve and the entropy density change, the only
term present which depends on the conductivity is the last one.

\subsection{Perpendicular shock}

Shock waves in a plasma permeated with a magnetic field show several
features, of which the most well known is related to the orientation of the
magnetic field with respect to the shock plane. Although perpendicular
shocks can be considered to be a special case of oblique shocks, it is
interesting to write the simplified expressions for the junction conditions
explicitly, and re-derive the adiabatic curve and the entropy density change
for this case.

\subsubsection{Hydrodynamical and electromagnetic junction conditions}

For perpendicular shocks $B_{n}=0$, so that the junction conditions now read

\begin{equation}
\left[ \rho v_{x}\right] =0,  \label{av}
\end{equation}

\begin{equation}
\left[ \rho v_{x}\left( \frac{1}{2}v^{2}+{\sf w}\right) +\frac{1}{4\pi }
v_{x}B_{t}^{2}-\frac{c^{2}}{16\pi ^{2}\sigma }\partial _{x}\left(
B^{2}\right) \right] =0,  \label{aw}
\end{equation}

\begin{equation}
\left[ \rho v_{x}^{2}+p+\frac{1}{8\pi }B_{t}^{2}\right] =0,  \label{ax}
\end{equation}

\begin{equation}
\left[ \rho \vec{v}_{t}v_{x}\right] =0\Rightarrow \vec{v}_{t}=\vec{v}_{t1},
\label{ay}
\end{equation}

\begin{equation}
\frac{c}{4\pi \sigma }\left[ \left( \partial _{x}\times \vec{B}_{t}\right) 
\right] -\frac{1}{c}\left[ v_{x}\vec{B}_{t}\right] =0.  \label{ba}
\end{equation}

\subsubsection{Adiabatic curve and entropy density change}

Proceeding as for an oblique shock and defining

\begin{equation}
\varepsilon ^{\ast }=\varepsilon +\frac{B_{t}^{2}V}{8\pi },\quad p^{\ast }=
p+\frac{1}{8\pi }B_{t}^{2},  \label{bi}
\end{equation}
we obtain

\begin{equation}
{\sf w}-{\sf w}_{1}-\frac{1}{2}\left[ p^{\ast }\right] \left( V+V_{1}\right)
+\frac{1}{4\pi }\left[ VB_{t}^{2}\right] -\frac{c^{2}}{16j\pi ^{2}\sigma }
\left[ \partial _{x}\left( B_{t}^{2}\right) \right] \left. =\right. 0
\label{be}
\end{equation}
and

\begin{equation}
\varepsilon ^{\ast }-\varepsilon _{1}^{\ast }+\frac{1}{2}\left( p^{\ast
}+p_{1}^{\ast }\right) \left( V-V_{1}\right) -\frac{c^{2}}{16j\pi ^{2}\sigma
}\partial _{x}\left( B_{t}^{2}\right) =0.  \label{bj}
\end{equation}
There is now only one term that depends on the dissipative properties of the
plasma, while for the oblique case, we had two such terms. The missing term
is related to the normal component of the magnetic field.

\section{Thickness of the shock wave}

The calculation of a general expression for the shock thickness is very
difficult, if not impossible. However for a perpendicular shock it is
possible to calculate the shock thickness exactly. We then have $B_{n}=0$, 
$\vec{v}_{t}=0$ and consider a coordinate system in which the only non-zero
component of the magnetic field is $B_{y}=B$, \cite{ecm-land}. In this case,
the equation $\vec{\nabla}.\vec{B}=0$ is satisfied identically. The
unidimensional ideal MHD equations are

\begin{equation}
\frac{\partial B}{\partial t}=\frac{\partial }{\partial x}\left(
v_{x}B\right) ,  \label{bs}
\end{equation}

\begin{equation}
\frac{\partial \rho }{\partial t}+\frac{\partial }{\partial x}\left(
v_{x}\rho \right) =0,  \label{bt}
\end{equation}

\begin{equation}
\frac{\partial v_{x}}{\partial t}+v_{x}\frac{\partial v_{x}}{\partial x}+
\frac{1}{8\pi \rho }\frac{\partial B^{2}}{\partial x}=-\frac{1}{\rho }
\frac{
\partial p}{\partial x}.  \label{bu}
\end{equation}
From the first two equations, it is easy to see that the ratio 
$B/\rho \equiv \beta $ satisfies the equation $\partial \beta /\partial
t+v_{x}\partial \beta /\partial x=0$ or $d\beta /dt=0$ \cite{ecm-land}.
Hence, if the fluid is homogeneous at some initial instant, so that $\beta
=const.$, then it will remain so at all subsequent times. Substituting 
$B=\rho \beta $ in the third equation, we obtain

\begin{equation}
\frac{\partial v_{x}}{\partial t}+v_{x}\frac{\partial v_{x}}{\partial x}=-
\frac{1}{\rho }\frac{\partial }{\partial x}\left[ p+
\frac{\beta ^{2}\rho ^{2}}{8\pi }\right]   \label{bv}
\end{equation}
Thus, the magnetic field has been eliminated from the equations. The
equation for the velocity field, Eq. (\ref{bv}), is formally identical to
that for the ideal fluid case, provided we define the\ `true pressure' as 
$p^{\ast }=p+\beta ^{2}\rho ^{2}/8\pi $. We can now proceed to evaluate the
thickness, following Ref. \cite{flm-land}. We write

\begin{equation}
\frac{\partial }{\partial t}\delta p^{\ast }-v_{s}^{\ast }
\frac{\partial }{\partial x}\delta p^{\ast }-\alpha _{p}^{\ast }
\delta p^{\ast }\frac{\partial }{\partial x}\delta p^{\ast }=
cL^{\ast }\frac{\partial ^{2}}{\partial x^{2}}\delta p^{\ast },  \label{bw}
\end{equation}
where $v_{s}^{\ast 2}\equiv \left( \partial p^{\ast }/\partial \rho \right)
_{s}$ and $L^{\ast }$ is the damping length, given by Eq. (\ref{a45}) of the
Appendix,

\begin{equation}
L^{\ast }=\frac{c}{8\pi \sigma }\frac{B_{0y}^{2}}{\left( B_{0y}^{2}+4\pi
\rho _{0}v_{s}^{2}\right) }.  \label{bx}
\end{equation}
From Ref. \cite{flm-land} we have

\begin{equation}
\alpha _{p}^{\ast }=\frac{1}{2}v_{s}^{\ast 3}\rho ^{2}\left[ \frac{\partial
^{2}}{\partial p^{\ast 2}}\left( \frac{1}{\rho }\right) \right] _{s}.
\label{by}
\end{equation}
Equation (\ref{bw}) can be solved using the procedure in Ref. \cite{flm-land}
, obtaining the thickness of the shock wave as

\begin{equation}
\delta ^{\ast }=\frac{4cL^{\ast }}{\alpha _{p}^{\ast }\left( p_{2}^{\ast
}-p_{1}^{\ast }\right) },  \label{bz}
\end{equation}
where $p_{2}^{\ast }$ and $p_{1}^{\ast }$ are the `true pressures' far
downstream and far upstream respectively. Equation (\ref{bz}) is
quantitatively valid for sufficiently small differences $\left( p_{2}^{\ast
}-p_{1}^{\ast }\right) $. However we can use it qualitatively to estimate
the order of magnitude of the thickness in cases where the difference 
$\left( p_{2}^{\ast }-p_{1}^{\ast }\right) $ is of the same order of
magnitude as $p_{2}^{\ast }$ and $p_{1}^{\ast }$ themselves. The velocity of
sound in the gas, $v_{s}$ (not $v_{s}^{\ast }$) is of the same order of
magnitude as the thermal velocity $v$. Let $\lambda $ be the mean free path
of the atoms in the plasma. Then from dimensional analysis, the electric
conductivity can be estimated as, $\sigma \sim \gamma v/\lambda \sim \gamma
v_{s}/\lambda $, where $\gamma $ takes into account anomalous effects and
can have a value $10^{-6}\leq \gamma \leq 1$ 
\footnote{We know that in accretion disks, protostars, galactic nuclei and 
neutron
X-ray sources, for example, the plasma cannot have ideal Spitzer values for
the conductivity and viscosity in order to obtain the observed accretion
rates. Therefore it is generally assumed that these quantities are highly
anomalous (due to turbulence, for example). Another example where the
assumption of anomalous resistivity is used is in the treatment of solar
flares, which are generally assumed to be due to magnetic reconnection. If
ideal Spitzer values are used for the plasma in solar flares, reconnection
times are $\sim 10^{6}$ times longer than the observed time scales for the
flares. In general, plasmas near shocks are expected to be highly anomalous
(i.e., $\gamma \ll 1$) due to turbulence.}.

In Eq. (\ref{bx}), we take $B_{0y}^{2}/\left( B_{0y}^{2}+4\pi \rho
_{0}v_{s}^{2}\right) \sim B_{0y}^{2}/p^{\ast }$ and in Eq. (\ref{by}) 
$\alpha _{p}^{\ast }\left( p_{2}^{\ast }-p_{1}^{\ast }\right) \sim
v_{s}^{\ast 2}\sim p^{\ast }/\rho $. Using these relations in Eq. (\ref{bz})
we obtain

\begin{equation}
\delta \sim c^{2}\frac{\rho ^{2}B_{0y}^{2}}{\gamma p^{\ast 2}p}\lambda .
\label{ca}
\end{equation}
The shock thickness is larger than the mean free path when $c^{2}\rho
^{2}B_{0y}^{2}\geq \gamma p^{\ast 2}p$. Let us first assume that 
$B_{0y}^{2}\geq p$. We then have $p^{\ast 2}\sim B_{0y}^{4}$ and 
$p\leq B_{0y}^{2}\leq c^{2}\rho ^{2}/\gamma p$. As a specific
numerical example, consider $\rho \sim 10^{2}$ gr/cm$^{3}$ and $T\sim 10^{8}$
K (characteristic parameters at the center of a massive star before
collapse). Assuming hydrogen gas, we have $n\sim 10^{26}$ cm$^{-3}$ and 
$p\sim nT\sim 10^{18}$erg/cm$^{3}$ (for iron nuclei, the pressure would be
two orders of magnitude smaller). For the above densities and pressures, the
shock thickness is larger than the mean free path of the particles if the
magnetic field is in the interval $10^{9}$ G $\leq B_{0y}\leq \gamma
^{-1/2}10^{14}$ G. If we now assume that $B_{0y}^{2}\leq p$, the shock
thickness is larger than the mean free path if $\gamma p^{3}/c^{2}\rho
^{2}\leq B_{0y}^{2}$. For the above parameters we have $\gamma
^{1/2}10^{5}$ G $\leq B_{0y}\leq 10^{10}$ G
\footnote{For $\gamma \sim 10^{-6}$, the magnetic field range is $10^{2}$ G 
$\leq B_{0y}\leq 10^{10}$ G. A magnetic field $B\geq 10^{2}$ G 
can be
easily be present at the center of a massive star. The magnetic field
increases in the collapse of the core of a massive star to $B\geq 10^{6}$
(for the collapse to a white dwarf) and to $B\geq 10^{9}$ (for the
collapse to a neutron star). Thus, at the center of a massive star, we may
expect that the magnetic field varie from $10^{2}$ G to $10^{10}$ G during
the collapse of its core and the start of a supernova explosion.}. If
neither of the two conditions above are fulfilled, the shock thickness is
smaller than the mean free path. This means that the MHD approach breaks
down and kinetic theory is needed to study the structure of the shock.

\section{Conclusions}

In this article, we extended the results of shock waves treated in ideal MHD
to the non-ideal case, in which the electrical conductivity is finite. We
considered Ohm's law in its simplest form, \cite{spitzer}, but took into
account phenomenologically (through the parameter $\gamma $) plasma effects
that can modify the classical Spitzer electrical conductivity (e.g.,
turbulence). The expressions for the adiabatic curve and the entropy density
change across the shock were generalized. Finally, we derived the expression
for the shock thickness for a finite conductivity in the case of a parallel
shock in strongly magnetized plasmas. The conditions that the ambient
magnetic field must satisfy for the thickness to be of the order of the
particle mean free path were estimated. We found that these conditions can
be fulfilled for the plasma expected in the origin of a supernova explosion,
\cite{star-1,star-2}. Extensions of the results presented in this paper,
using a more general Ohm's law, as well as to relativistic shocks, are
presently under investigation.

\begin{acknowledgments}
This work was partially supported by the Brazilian
financing agency FAPESP (00/06770-2). A.K. acknowledges the FAPESP
fellowship (01/07748-3). R. O. acknowledges partial support from the
Brazilian financing agency CNPq (300414/82-0).
\end{acknowledgments}

\section{Appendix}
In this appendix, we sketch the derivation of the expression 
for $L^{\ast }$
, the damping length used to calculate the shock thickness. Neglecting the
displacement current (which is a good approximation in non-relativistic
electrodynamics), the evolution equation for the magnetic field in a medium
with electrical conductivity $\sigma $ moving with a velocity $\vec{v}$ is

\begin{equation}
\frac{\partial \vec{B}}{\partial t}-\vec{\nabla}\times \left( \vec{v}\times 
\vec{B}\right) =\frac{c^{2}}{4\pi \sigma }\nabla ^{2}\vec{B}.  \label{a5}
\end{equation}
Adding the equations for the fluid,  which we assume has neither viscosity
nor thermal conduction, we have

\begin{equation}
\frac{\partial \rho }{\partial t}+\vec{\nabla}.\left( \rho \vec{v}\right) =0,
\label{a6}
\end{equation}

\begin{equation}
\frac{\partial \vec{v}}{\partial t}+\left( \vec{v}.\vec{\nabla}\right) 
\vec{v}=-\frac{1}{\rho }\vec{\nabla}p-\frac{1}{4\pi \rho }\vec{B}\times 
\left( \vec{\nabla}\times \vec{B}\right) .  \label{a7}
\end{equation}

\subsection{Hydromagnetic waves}

Let us assume that $B=\vec{B}_{0}+\vec{b}$, $\rho =\rho _{0}+\delta \rho $, 
$p=p_{0}+\delta p$, and $\vec{v}=\delta \vec{v}$. Replacing these terms in
the above equations, keeping only terms to first order in the perturbations,
expanding the density in powers of the perturbation in the pressure (i.e., 
$\delta \rho =\delta p/v_{s}^{2}+\left( \partial ^{2}\rho /\partial
p^{2}\right) _{s}\delta p^{2}$, where $v_{s}$ is the sound velocity of the
medium) and taking the Fourier transform of the equations, we obtain

\begin{equation}
-\omega \delta p_{0}+v_{s}^{2}\rho _{0}\vec{k}.\delta \vec{v}_{0}=0,
\label{a23}
\end{equation}

\begin{equation}
-\left( \omega +i\frac{c^{2}}{4\pi \sigma }k^{2}\right) \vec{b}_{0}=
\vec{k}\times \left[ \delta \vec{v}_{0}\times \vec{B}_{0}\right] ,  
\label{a24}
\end{equation}

\begin{equation}
-\omega \delta \vec{v}_{0}=-\frac{\vec{k}}{\rho _{0}}\delta p_{0}-
\frac{1}{4\pi \rho _{0}}\vec{B}_{0}\times \left( \vec{k}\times \vec{b}_{0}
\right) .
\label{a25}
\end{equation}
From Eq. (\ref{a23}), we find $\delta p_{0}=v_{s}^{2}\rho _{0}\left( 
\vec{k}.\delta \vec{v}_{0}\right) /\omega $ and using this in Eq. 
(\ref{a25} ), we obtain

\begin{equation}
-\delta \vec{v}_{0}=-\frac{\vec{k}}{\omega }v_{s}^{2}\left( 
\frac{\vec{k}}{\omega }.\delta \vec{v}_{0}\right) -
\frac{1}{4\pi \rho _{0}}\vec{B}_{0}\times \left( 
\frac{\vec{k}}{\omega }\times \vec{b}_{0}\right) .
\label{a26}
\end{equation}

We define the scalar phase velocity as $u=\omega /k$, assuming that $\vec{k}$
is along the $x$-axis, (i.e., $\vec{k}=k\check{x}$) and that $\vec{B}_{0}$
is in the $x-y$ plane. Writing the previous equations in its components, we
have

\begin{equation}
\delta p_{0}=\frac{v_{s}^{2}\rho _{0}}{u_{A}}\delta v_{0x},  \label{a27}
\end{equation}

\begin{equation}
\left( u-\frac{v_{s}^{2}}{u}\right) \delta v_{0x}=\frac{1}{4\pi \rho _{0}}
b_{0y}B_{0y},  \label{a28}
\end{equation}

\begin{equation}
u\delta v_{0y}=-\frac{1}{4\pi \rho _{0}}b_{0y}B_{0x},  \label{a29}
\end{equation}

\begin{equation}
\left( u+i\frac{c^{2}}{4\pi \sigma }k\right) b_{0y}=\delta
v_{0x}B_{0y}-\delta v_{0y}B_{0x},  \label{a30}
\end{equation}

\begin{equation}
u\delta v_{0z}=-\frac{1}{4\pi \rho _{0}}b_{0z}B_{0x},  \label{a31}
\end{equation}

\begin{equation}
\left( u+i\frac{c^{2}}{4\pi \sigma }k\right) b_{0z}=-\delta v_{0z}B_{0x}.
\label{a32}
\end{equation}

\subsection{Generalized Alfven waves}

Using Eqs. (\ref{a31}) and (\ref{a32}) we obtain the compatibility
relationship

\begin{equation}
u^{2}+i\frac{c^{2}k}{4\pi \sigma }u-\frac{B_{0x}^{2}}{4\pi \rho _{0}}=0,
\label{a33}
\end{equation}
from which we obtain

\begin{equation}
u=\pm \frac{1}{2}\sqrt{\frac{B_{0x}^{2}}{\pi \rho _{0}}-
\frac{c^{4}k^{2}}{16\pi ^{2}\sigma ^{2}}}-i\frac{c^{2}k}{8\pi \sigma }.  
\label{a34}
\end{equation}
From Eq. (\ref{a34}), the phase velocity is a complex number if $\sigma $ is
finite. Rewriting $u_{A}$ in terms of $\omega $, we obtain the dispersion
relationship:

\begin{equation}
\omega =\frac{1}{2}\frac{B_{0x}k}{\sqrt{\pi \rho _{0}}}\sqrt{1-\frac{\rho
_{0}}{B_{0x}^{2}}\frac{c^{4}k^{2}}{16\pi \sigma ^{2}}}-i
\frac{c^{2}k^{2}}{8\pi \sigma }.  \label{a35}
\end{equation}
We take the plus sign in eq. (\ref{a34}) since the frequency is a positive
quantity. The fact that the imaginary part is non-linear in $k$ means that
the Alfven waves are damped and dissipated as a function of $k$. For 
$\sigma \rightarrow \infty $ we recover the known dispersion relationship 
for ideal MHD.

Assuming that the second term in the square root in Eq. (\ref{a35}) is much
smaller than unity, the group velocity is

\begin{equation}
v_{A}=\frac{\partial \omega }{\partial k}\simeq \frac{B_{0x}}{\sqrt{4\pi
\rho _{0}}}-i\frac{c^{2}k}{4\pi \sigma }.  \label{a37}
\end{equation}
When the electrical conductivity is infinite, we recover the known ideal MHD
result, $v_{AI}=B_{0x}/2\sqrt{\pi \rho _{0}}$.

\subsection{Generalized magnetosonic waves}

From Eqs. (\ref{a28}), (\ref{a29}) and (\ref{a30}), we obtain the
generalized dispersion relationship for magnetosonic waves,

\begin{equation}
\omega ^{4}-\left( \frac{B_{0}^{2}}{4\pi \rho _{0}}+v_{s}^{2}\right)
k^{2}\omega ^{2}+\frac{B_{0x}^{2}}{4\pi \rho _{0}}v_{s}^{2}k^{4}+i
\frac{c^{2}k^{2}}{4\pi \sigma }\omega ^{3}-i
\frac{c^{2}v_{s}^{2}k^{4}}{4\pi \sigma 
}\omega =0.  \label{a38}
\end{equation}
This relationship can be inverted to obtain $\omega =\omega \left( k\right) $
. However, for our purposes, it suffices to consider the dissipative terms
as a correction to the ideal dispersion relationship,

\begin{equation}
\omega _{I}^{2}=\frac{1}{2}\left[ \left( \frac{B_{0}^{2}}{4\pi \rho _{0}}
+v_{s}^{2}\right) \pm \sqrt{\left( \frac{B_{0}^{2}}{4\pi \rho _{0}}
+v_{s}^{2}\right) ^{2}-\frac{B_{0x}^{2}}{\pi \rho _{0}}v_{s}^{2}}\right]
k^{2}\equiv v_{g0}^{2}k^{2}.  \label{a40}
\end{equation}
The plus sign corresponds to {\it fast magnetosonic} waves, while the minus
sign to {\it slow magnetosonic} waves. Replacing $\omega _{I}$ in the last
two terms of Eq. (\ref{a38}), we obtain

\begin{equation}
\omega ^{4}-\left( \frac{B_{0}^{2}}{4\pi \rho _{0}}+v_{s}^{2}\right)
k^{2}\omega ^{2}+\frac{B_{0x}^{2}}{4\pi \rho _{0}}v_{s}^{2}k^{4}+i
\frac{c^{2}v_{g0}}{4\pi \sigma }\left( v_{g0}^{2}-v_{s}^{2}\right) 
k^{5}\simeq 0.
\label{a41}
\end{equation}
If the electrical conductivity is large (but not infinite), we have

\begin{equation}
\omega ^{2}\simeq \omega _{I}^{2}\mp \frac{i}{4}\frac{c^{2}v_{g0}}{\pi
\sigma }\frac{\left( v_{g0}^{2}-v_{s}^{2}\right) }{\left(
v_{B0}^{4}-v_{B0x}^{2}v_{s}^{2}\right) ^{1/2}}k^{3},  \label{a42}
\end{equation}
where $v_{B0}^{2}=\left( B_{0}^{2}/4\pi \rho _{0}+v_{s}^{2}\right) $ and 
$v_{B0x}^{2}=B_{0x}^{2}/\pi \rho _{0}$. We thus have

\begin{equation}
\omega \simeq v_{g0}k-icL^{\ast }k^{2},  \label{a43}
\end{equation}
with

\begin{equation}
L^{\ast }=\frac{c}{8\pi \sigma }\frac{\left( v_{g0}^{2}-
v_{s}^{2}\right) }{\left( v_{B0}^{4}-v_{B0x}^{2}v_{s}^{2}\right) ^{1/2}},  
\label{a44}
\end{equation}
where $L^{\ast }$ is the damping length. For a perpendicular shock, (i.e., 
$B_{0x}=0$), we have

\begin{equation}
L^{\ast }=\frac{c}{8\pi \sigma }\frac{B_{0t}^{2}}{\left( B_{0t}^{2}+4\pi
\rho _{0}v_{s}^{2}\right) }.  \label{a45}
\end{equation}

\end{document}